\documentclass{aa}
\usepackage{txfonts}
\usepackage{graphicx}
\usepackage[normalem]{ulem}
\begin{document}
\title{Kinematics of Nearby Subdwarf Stars}
\author{M.I. Arifyanto \and  B. Fuchs \and H. Jahrei{\ss} \and R. Wielen}
\offprints{M. Ikbal Arifyanto,
\email{arifyanto@ari.uni-heidelberg.de}} \institute{Astronomisches
Rechen-Institut, M\"onchhofstra{\ss}e 12-14, D-69120 Heidelberg,
Germany}
\date{Received  / Accepted  }

\abstract{We present an analysis of the space motions of 742
subdwarf stars based on the sample of Carney et
al.(\cite{carney94}). Hipparcos parallaxes, TYC2+HIP proper
motions and Tycho2 proper motions were combined with radial
velocities and metallicities from CLLA. The kinematical behavior
is discussed in particular in relation to their metallicities. The
majority of these sample stars have metal abundances of [Fe/H] $>-1$ and represent the thick disk population. The halo component,
with [Fe/H] $<-1.6$, is characterized by a low mean rotation
velocity and a radially elongated velocity ellipsoid. In the
intermediate metallicity range ($-1.6\leq$ [Fe/H] $\leq-1$), we
find a significant number of subdwarfs with disklike
kinematics. We interpret this population of stars as a metal-weak
thick disk (MWTD) population.
\keywords{stars: subdwarfs -- Galaxy: kinematics and dynamics -- Galaxy: solar neighbourhood}}
\maketitle
\section{Introduction}

Studies of the kinematics of various stellar populations in the Galaxy,
 in particular the thick disk and the halo, have long been limited by
the availability of large samples of stars with measurements of
proper motions, radial velocities, distances, and metallicities.
Such data are required in order to constrain plausible
scenarios for the formation and evolution of the Milky Way.
Samples of nearby subdwarf stars with high space motions provide
an observationally convenient probe of the structure of the Galaxy.
The large, proper-motion-selected, stellar samples of Carney,
Latham, Laird, and Aguilar (\cite{carney94},
hereafter CLLA) have proved particularly valuable for studying the
kinematics and chemical abundances within a few kiloparsecs of
the Sun.

The correlation between kinematics and metallicity gives useful
information for formulating theories of galactic structure.
Differences in chemistry and space velocities are crucial in
defining the different populations within the Galaxy and inferring
their origins. Relevant studies of the kinematical behavior of
stars ,in particular in relation to their metallicities ,were presented 
by e.g. Morrison, Flynn, \& Freeman (\cite{morisson}, hereafter MFF) using
a sample of K giants whose metallicities are measured using the DDO
photometric system, Nissen \& Schuster (\cite{nissen}) using late F and G
dwarfs and subgiants, Chiba \& Yoshii (\cite{chiba98}) using red giants and
RR Lyrae stars, Martin \& Morrison (\cite{martin}) with a sample of nearby
RR Lyrae stars and Chiba \& Beers (\cite{chiba00}) using 1203 metal-poor
solar-neighborhood stars.

The Galactic halo is characterized by a roughly spherical space
distribution with close to zero net rotation. Its stars are metal
poor, with a peak metallicity at [Fe/H] $=-1.6$ (Laird et al.
{\cite{laird}}). The halo population in the solar neighborhood is not purely
a relic of a monolithic, ``rapid'' collapse (Carney et al. {\cite{carney96}}).
There have been several suggestions of a two-component halo, with
a flattened component in the inner halo and a more spherical outer
halo (Sommer-Larsen \& Zhen {\cite{slarsen}}; Carney et al. {\cite{carney96}}).

The Galactic thick disk is the kinematically hottest portion of
the disk of the Galaxy, with a scale height of 1.0 to 1.5 kpc and
rotates with a velocity of about 170 km s$^{-1}$ (Gilmore, Wyse,
\& Kuijken {\cite{gilmore89}}). The thick disk is usually considered to be dominated by stars 
in the range [Fe/H] $> -1$ (Freeman {\cite{freeman}}), peaking at about
[Fe/H] $=-0.5$ (Carney et al. {\cite{carney89}}). Many workers have claimed the
existence of a metal weak tail of the thick disk component in the
range $-1.6 \le$ [Fe/H] $\le-1$ (MFF; Beers \& Sommer-Larsen {\cite{beers95}};
Chiba \& Beers {\cite{chiba00}}). MFF found a fraction of $72$\% of the
stars in this metallicity range in a
"metal-weak thick disk" (MWTD), rotating rapidly at $V_\mathrm{rot} \approx 170$ km s$^-1$.
Another large fraction of MWTD was found also by Beers \&
Sommer-Larsen (\cite{beers95}). Their MWTD, rotating at $V_\mathrm{rot} \approx 195$ km
s$^-1$, accounts for about 60 \% of the stars in the range $-1.6
\le$ [Fe/H] $\le-1$ in the solar neighborhood, and it possesses an
extremely metal-weak tail down to [Fe/H] $\le -2$. Chiba \& Beers
(\cite{chiba00}) estimated the fraction of MWTD at about 30 \% of the
metal-poor stars in the abundance range $-1.7 \le$ [Fe/H] $\le-1$,
which is smaller than the fraction derived by MFF and Beers \&
Sommer-Larsen (\cite{beers95}), but larger by $\sim 10 \%$ than the result of Chiba \&
Yoshii (\cite{chiba98}) using solar neighborhood red giants
and RR Lyrae stars.

The investigation of thick-disk and halo kinematics may only be
applicable to a specific place in the Galaxy and may have fine
structure of the velocity distribution smoothed out by the
velocity resolution of the study
(Martin \& Morrison {\cite{martin}}). Here, we study the kinematics
of solar neighborhood subdwarf stars based on the sample of high
proper motion stars by CLLA. CLLA have measured photometric
parallaxes, radial velocities, and metallicities of mainly A
to early G stars, many late G and some early K stars in the {\it
Lowell Proper Motion Catalogue}. In total the CLLA sample contains
$1464$ stars. In their paper there are listed 1269
stars with kinematical parameters and 1261 stars with metallicity
parameters, and there are
1447 stars with radial velocities in their catalog. The radial
velocity precision of their sample lies in the range of $0.4$ to
$1.3$ km s$^{-1}$. About 15\% of their sample are binaries or
multiple systems. The typical accuracy of the metallicities was
estimated to be $\pm 0.13$ dex.

\begin{figure}
\resizebox{\hsize}{!}{\includegraphics{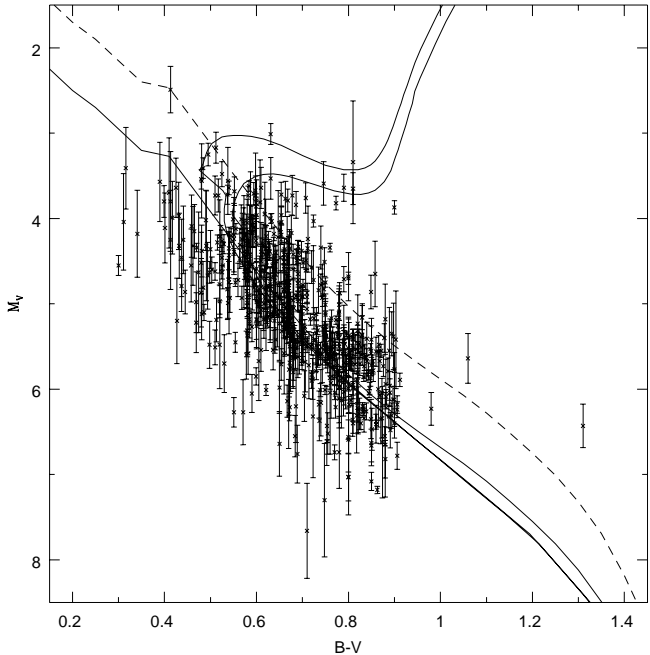}}
\caption{Color-Magnitude diagram for all identified CLLA stars.
Hipparcos parallaxes were used to determine $M_{V}$ and its
standard error. The full lines indicate the zero age main sequence
(ZAMS) and main sequences of the old open clusters M 67 and NGC
188. The dashed line is the ZAMS shifted upward by $\Delta
M_{V}=0.8$ mag., used to remove the contamination by subgiants and
giants.} \label{cmd}
\end{figure}

\begin{figure}
\resizebox{\hsize}{!}{\includegraphics{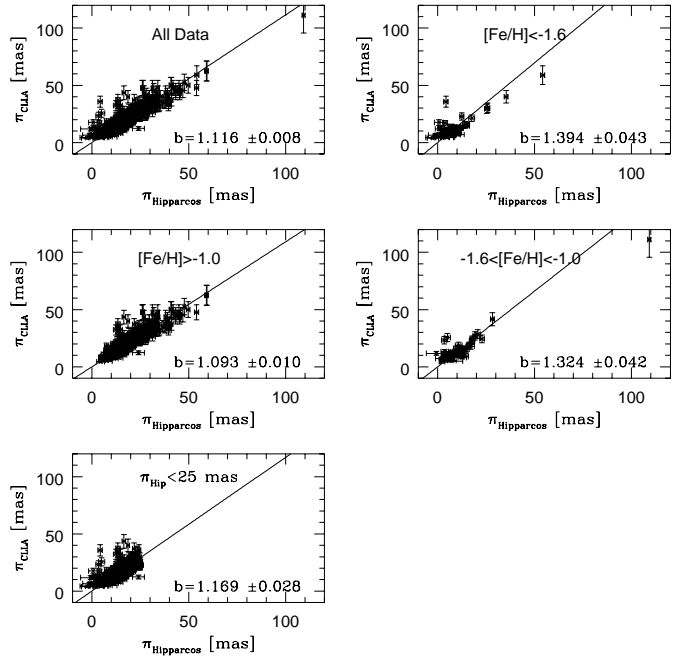}}
\caption{Hipparcos trigonometric parallaxes versus the photometric
parallaxes of CLLA, for 539 stars (top-left) and for different
metallicity cuts : [Fe/H] $<-1.6$ (top-right), [Fe/H] $>-1$
(center-left), $-1.6 \le$ [Fe/H] $\le -1$ (center-right), and for stars at large
distances, $\pi_\mathrm{Hip} < 25$ mas (bottom-left). The full line is a linear
fit to the data.}
\label{fig2}
\end{figure}

The photometric parallax of CLLA was replaced in our study by using
the high precision parallax of Hipparcos catalogue. We used the
{\it Astrometric Catalog TYC2+HIP} (Wielen et al. {\cite{arihip01}}) for the
proper motions of stars with Hipparcos parallaxes. This catalogue
is derived from a combination of the Hipparcos Catalogue with
proper motions given in the Tycho-2 catalogue with direct
solutions (Wielen et al. {\cite{arihip01}}) and previous earthbound measurements. We still use the high precision
radial velocities and metallicities of the CLLA catalogue for our
study. Previous work using Hipparcos subdwarfs was done by
Reid (\cite{reid}) and Fuchs, Jahrei{\ss}  and Wielen (\cite{fuchs}; hereafter
FJW). In this previous work we discussed the kinematical behaviour of the 560
subdwarfs for which improved parallaxes and proper motions were obtained by
Hipparcos, in relation to their metallicities. In the present paper we increase
the size of the sample considerably by applying a correction to the photometric
CLLA distances determined using stars with Hipparcos parallaxes.

In Sect. 2 we describe the selection of the sample and the photometric correction.
In the next section we determine the
kinematical properties, i.e. the 3D velocities with respect to the
Sun, and discuss the kinematical behavior of the subdwarfs, in
particular in relation to their metallicities. Our conclusions are
summarized in Sect. 4.

\section{Data}

The CLLA data set of 1447 stars has been cross-identified with the
Astrometric Catalogue TYC2+HIP (Wielen et al. {\cite{arihip01}}) and we found
545 stars in common. About 700 CLLA stars did not appear
in the TYC2+HIP catalogue, but were then cross-identified with the
Tycho-2 Catalogue (H\o g et al. {\cite{tycho2}}), and we found 259 subdwarfs
with Tycho-2 proper motions. The proper motion accuracy of Tycho 2,
derived from a comparison with the
Astrographic Catalogue and 143 other ground-based astrometric
catalogues, is about $2.5$ mas yr$^{-1}$. CLLA used Luyten's
NLTT proper motions for the
calculation of the space velocity components. These proper motions
have typical errors of 20 to 25 mas yr$^{-1}$. Therefore, TYC2+HIP and
Tycho-2 proper motions provide an enormous improvement in the
accuracy of the tangential velocities.

\subsection{Color Magnitude Diagram}
Fig. 1 shows the color magnitude diagram for all 545 identified
CLLA stars. The absolute magnitudes and
their standard errors are based on the Hipparcos parallaxes and
their errors. The $B-V$ colors were taken from the Hipparcos
catalogue. Some stars for which no distance was given were already
recognized by CLLA as subgiants. However, we can see clearly from
the CM-diagram in Fig. 1 that there is still contamination by previously
undetected subgiants and giants. To avoid these, we have
removed all stars lying above a line in the CM-diagram defined by
the zero age main sequence of stars with solar metallicity shifted
upward by $\Delta M_\mathrm{V}=0.8$ mag. About 8\% contaminating stars were
eliminated in this way.

\subsection{Test of Photometric Distances}

We determined the overall correction of the photometric
distance scale of CLLA by analyzing the parallax difference. There
are 539 CLLA stars which have both photometric and trigonometric
parallaxes in our sample (CLLA-TYC2+HIP). We compared the
Hipparcos parallax with the photometric parallax of CLLA (as shown in Fig.
2). The error bars represent both Hipparcos and CLLA parallax
errors. A typical error in the absolute magnitude of CLLA stars of
$\sigma_{M_\mathrm{V}}=0.3$ mag is assumed. We used a least $\chi^{2}$
method applicable when the data have errors in both coordinates. The
$\chi^{2}$-function is chosen according to Press et al.(\cite{press})
\begin{equation}\label{chi}
    \chi^{2}=\sum ^{N}_{i=1} \frac{(y_{i}-bx_{i})^{2}}{\sigma^{2}_{y_{i}}+b^{2}\sigma^{2}_{y_{i}}}
\end{equation}

\begin{figure}
\resizebox{\hsize}{!}{\includegraphics{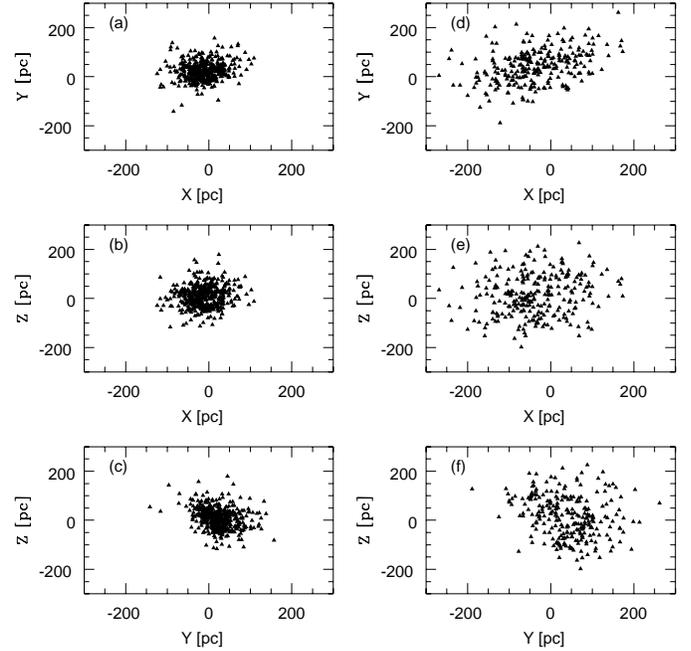}}
\caption{Spatial distribution of the samples CLLA-TYC2+HIP (a to c) and CLLA-Tycho-2 (d to f)
, respectively. X points towards the Galactic center, Y in the direction of galactic rotation
and Z towards the Galactic north pole.} \label{position}
\end{figure}

The slope of the regression $b$ derived from our 539 subdwarfs, is
$b=1.116 \pm 0.008$. For comparison, a sample with
$\pi_\mathrm{Hip}\le 25$ mas leads to a larger correction of
$b=1.169 \pm 0.028$. We also tried cuts in the metallicties, with
the result that for more metal-poor stars a larger correction was
needed. For stars with [Fe/H] $>-1$, which are dominated by thick
disk stars, $-1.6 \le$ [Fe/H] $\le -1$, and extreme metal-poor stars
with [Fe/H] $<-1.6$ we find slopes of the regression line $b=1.093
\pm 0.010$, $b=1.324 \pm 0.042$ and $b=1.394 \pm 0.043$,
respectively. Jahrei\ss, Fuchs and Wielen (\cite{jahreiss}) and
FJW have found similar corrections using a smaller sample of
subdwarfs. The data points at the upper right corner of the
first plot (All Data) and the fourth plot ($-1.6 \le$ [Fe/H] $\le
-1$) represent the star HIP $57939$, which is the nearest star
in our sample with $\pi_{\mathrm{Hip}}=109$ mas. No significant changes in
the slopes of both plots (less than $1\sigma$) are found, if we
omit HIP $57939$ when calculating the slopes.

All these corrections are used to calibrate CLLA
photometric parallaxes in our sample with no Hipparcos
parallaxes and about 35 stars with low
accuracy Hipparcos parallaxes
i.e. $\pi_\mathrm{Hip} < 5$
mas and $\pi_\mathrm{Hip}/\sigma_{\pi_\mathrm{Hip}} < 3$ mas.

\begin{figure}
\resizebox{\hsize}{!}{\includegraphics{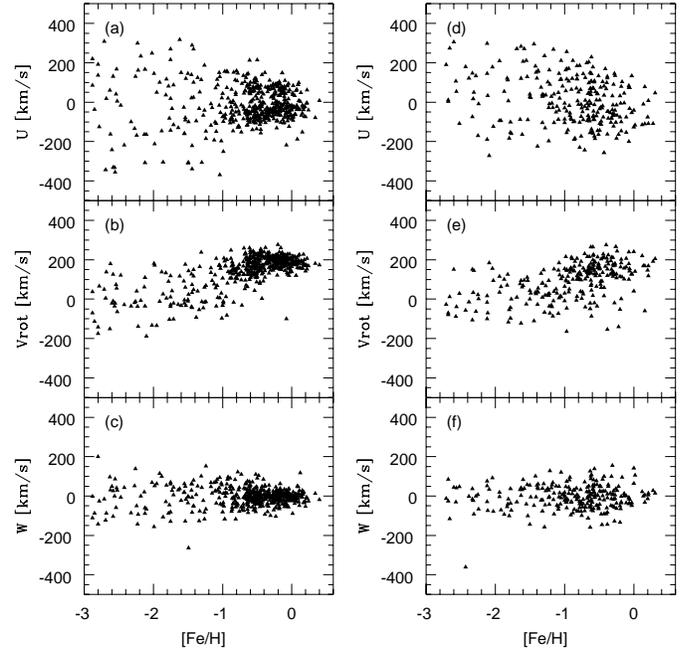}}
\caption{Space velocity components $(U, V_\mathrm{rot}, W)$
versus metallicity [Fe/H] of
the samples CLLA-TYC2+HIP (a to c) and CLLA-Tycho-2 (d to f) ,
respectively.} \label{velocity}
\end{figure}

\section{Kinematical Properties}

Using the parallaxes supplied by the Hipparcos catalogue, proper
motions by the TYC2+HIP and Tycho-2 catalogue, and radial
velocities given in CLLA, the space velocity components $U,V,$ and
$W$, which are directed to the Galactic center, direction of galactic rotation,
and north galactic pole, respectively, have been
calculated with respect to the Sun and then reduced to the LSR
(local standard of rest).  For the latter Delhaye's (\cite{delhaye})
values $+9, +12, +7$
km s$^{-1}$ were adopted for $U_{\sun},V_{\sun}, W_{\sun}$,
 respectively. Finally, the velocity components were transformed
onto a frame rotating with circular velocity $V_\mathrm{circ}=-220$ km
s$^{-1}$ relative to the LSR, i.e. the expected rest frame of our
Galaxy (e.g. Wielen {\cite{wielen}}). The rotational velocity is defined
as V$_\mathrm{rot}=V-V_\mathrm{circ}$.

\begin{table*}
\caption{Mean Velocities and Velocity Dispersion of the Sample Stars}
\begin{center}
\vspace{0.3cm}
\begin{tabular}{l r r @{$\pm$}l r @{$\pm$}l r @{$\pm$}l r @{$\pm$}l r @{$\pm$}l r @{$\pm$}l}
  \hline
  \hline
    & N &
  \multicolumn{2}{c}{$\langle U \rangle$}        & \multicolumn{2}{c}{$\langle V \rangle$}        & \multicolumn{2}{c}{$\langle W \rangle$}      &
  \multicolumn{2}{c}{$\sigma_\mathrm{U}$} & \multicolumn{2}{c}{$\sigma_\mathrm{V}$} & \multicolumn{2}{c}{$\sigma_\mathrm{W}$} \\
  dex&&\multicolumn{2}{c}{}&\multicolumn{2}{c}{}&\multicolumn{2}{c}{km s$^{-1}$}&\multicolumn{2}{c}{}&\multicolumn{2}{c}{}&\multicolumn{2}{c}{}\\
  \hline
  \bf{CLLA-TYC2+HIP}& &\multicolumn{1}{c}{} &\multicolumn{1}{c}{} &\multicolumn{1}{c}{} &
  \multicolumn{1}{c}{} &\multicolumn{1}{c}{} &\multicolumn{1}{c}{}  \\
  ${\rm{[Fe/H]}}>-1.0$ & $381$ & $-11$&$4$ & $-44$&$3$ & $-4$&$2$ & $74$&$2$ & $50$&$1$ & $37$&$1$ \\
  $-1.6 \le $ [Fe/H] $ \le-1.0$ & $49$ & $-35$&$23$ & $-163$&$12$ & $-1$&$11$ & $158$&$11$ & $85$&$6$ & $74$&$5$ \\
  ${\rm{[Fe/H]}}<-1.6$ & $53$ & $-5$&$26$ & $-219$&$13$ & $-1$&$13$ & $189$&$13$ & $97$&$7$ & $98$&$7$\\
  \hline
\bf{CLLA-Tycho2}& & \multicolumn{1}{c}{}& \multicolumn{1}{c}{}&
\multicolumn{1}{c}{}&
\multicolumn{1}{c}{} &\multicolumn{1}{c}{} &\multicolumn{1}{c}{}  \\
  ${\rm{[Fe/H]}}>-1.0$ & $169$ & $-9$&$8$ & $-91$&$6$ & $-5$&$4$ & $110$&$4$ & $81$&$3$ & $58$&$2$ \\
  $-1.6 \le $ [Fe/H] $ \le -1.0$ & $50$ & $51$&$17$ & $-180$&$10$ & $4$&$9$ & $121$&$9$ & $68$&$5$ & $61$&$4$ \\
  ${\rm{[Fe/H]}}<-1.6$ & $40$ & $36$&$25$ & $-205$&$14$ & $-19$&$12$ & $157$&$12$ & $87$&$7$ & $77$&$6$\\
  \hline
\end{tabular}\label{tab1}
\end{center}
\end{table*}

\begin{figure}
\resizebox{\hsize}{!}{\includegraphics{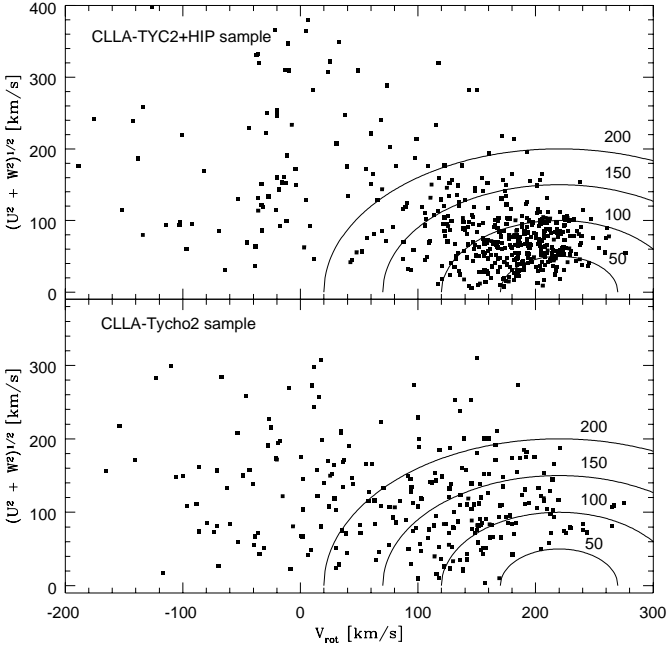}}
\caption{Toomre diagram: $(U^{2}+W^{2})^{1/2}$ versus $V_\mathrm{rot}$ distributions of CLLA-TYC2+HIP and CLLA-Tycho2 samples. The solid lines represent total velocities of $50, 100, 150$ and $200$ km s$^{-1}$, respectively.} \label{toomre}
\end{figure}

Figure {\ref{velocity}} shows the U, V$_\mathrm{rot}$, and W
velocities of the samples CLLA-TYC2+HIP and CLLA-Tycho2 as scatter
plots. The U-distribution indicates that the present sample
was kinematically selected. The CLLA Catalog is based on a proper
motion catalog so that stars with small tangential velocities are
missing. We can see clearly that for small U-values the diagrams
are sparsely populated. This is also seen in the V-velocities. The
stars with metallicities [Fe/H] $>-1$ lag on the average by about
$40$ km s$^{-1}$, i.e. are thick disk stars. The old thin disk
stars are missing (cf. also Fig.\ref{toomre}). In the W-velocities
no kinematical bias is visible; it is apparently lost in
projection. However, since we are mainly interested in the
kinematics of the halo stars, this bias is of no consequence in
the present context. It is evident from this figure that metal-poor stars with [Fe/H] $<-1$ have larger random motions compared
with metal-rich ones [Fe/H] $>-1$. This shows that the kinematic
properties change rather abruptly at [Fe/H] $\approx \: -1$ to $-2$, which is
probable the transition region from halo to disk component (Ryan
\& Norris, {\cite{ryan}} and Chiba \& Yoshii, {\cite{chiba98}}).

The mean motion with respect to the LSR and velocity dispersions were
calculated for different groups in [Fe/H]. The results are
presented in Table \ref{tab1}. The most metal-deficient stars in
the samples, more metal poor than [Fe/H] $=-1.6$, are
dominated by members of the halo population. These stars exhibit a
radially elongated velocity ellipsoid $(\sigma_\mathrm{U}, \sigma_\mathrm{V},
\sigma_\mathrm{W})=(189 \pm 13, 97 \pm 7, 98 \pm 7)$ and $(157 \pm 12, 87
\pm 7, 77 \pm 6)$ km s$^{-1}$ and show no net rotation, $\langle V_\mathrm{rot} \rangle =1 \pm 13$ and $15\pm 14$ km s$^{-1}$ for the CLLA-TYC2+HIP
and CLLA-Tycho2
samples, respectively, which are in good agreement with RR Lyrae
kinematics of Martin \& Morrison (\cite{martin}) and Layden et al. (\cite{layden}).
Chiba \& Beers (\cite{chiba00}) found a lower velocity dispersion in the
U-direction, $(\sigma_\mathrm{U}, \sigma_\mathrm{V}, \sigma_\mathrm{W})
=(141 \pm 11, 106
\pm 9, 94 \pm 8)$ km s$^{-1}$ from their 1203 non-kinematically selected
stars.

The velocity dispersion components of the sample in the
more metal-rich abundance ranges decrease as the contribution of
the thick disk component progressively increases. In particular,
for [Fe/H] $> -1.0$ the contribution of the halo component
is expected to be negligible. Our CLLA+TYC2+HIP sample in this
metallicity range has velocity dispersions $(\sigma_\mathrm{U},
\sigma_\mathrm{V}, \sigma_\mathrm{W})=(74 \pm 2, 50 \pm 1, 37 \pm 1)$ with
$V_\mathrm{rot}= 176$ km s$^{-1}$, which is in agreement with thick disk
samples of Martin \& Morrison (\cite{martin}) of RR Lyrae stars and
Chiba \& Beers (\cite{chiba00}) of solar-neighborhood stars.

Our CLLA-Tycho2 sample is more sparsely distributed than the
CLLA-TYC2+HIP sample in the metallicity range [Fe/H] $>-1$. To understand
this, we note that the CLLA-Tycho2 sample was drawn
from the CLLA stars that are not in the Hipparcos
catalogue. This might imply that mainly CLLA stars with
magnitudes brighter than $10.5$ mag fall in our CLLA-TYC2+HIP
sample and stars with magnitudes fainter than $10.5$ mag are
in the CLLA-Tycho2 sample. Stars with fainter apparent magnitudes are at
larger distances and
velocities compared to the Hipparcos stars. The minimum distances
for each sample are $17$ and $40$ pc for CLLA-TYC2+HIP and
CLLA-Tycho2, respectively. Figure {\ref{position}}, where we plot
the spatial distributions in X, Y and Z shows this clearly. We can
find the minimum tangential velocity using the minimum distances
and mean proper motions for both samples, using

\begin{eqnarray}
  V_\mathrm{T\:min} &=& 4.74 \frac{\langle \mu \rangle}{1000} d_\mathrm{min}
  \label{tangential}
\end{eqnarray}

where $\langle \mu \rangle$, $d_\mathrm{min}$, and $V_\mathrm{T\:min}$
denote mean proper motions in mas yr$^{-1}$, minimum distances in
parsecs and minimum tangential
velocities in km s$^{-1}$ for each subdwarf sample. We found for
the CLLA-Tycho2 sample $V_\mathrm{T\:min} >  50$ km s$^{-1}$,
which might explain why there are comparatively few thick disk stars in this sample
(cf. Fig.{\ref{toomre}}).

\subsection{$V_{rot}$ distributions of subdwarfs}

\begin{figure}
\resizebox{\hsize}{!}{\includegraphics{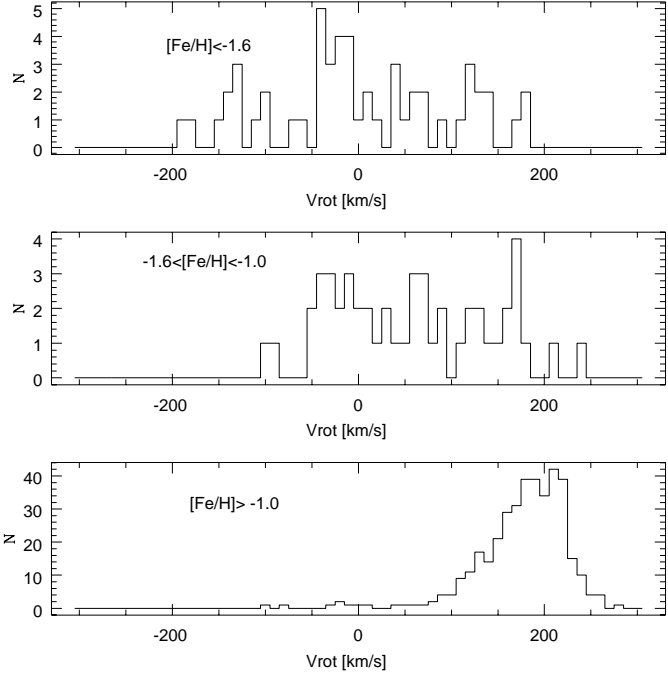}}
\caption{Rotational velocity (V$_\mathrm{rot}$) distributions of
sample CLLA-TYC2+HIP grouped according to their metallicities. The
velocities are reduced to the local standard of rest.}
\label{histo_ari}
\end{figure}

\begin{figure}
\resizebox{\hsize}{!}{\includegraphics{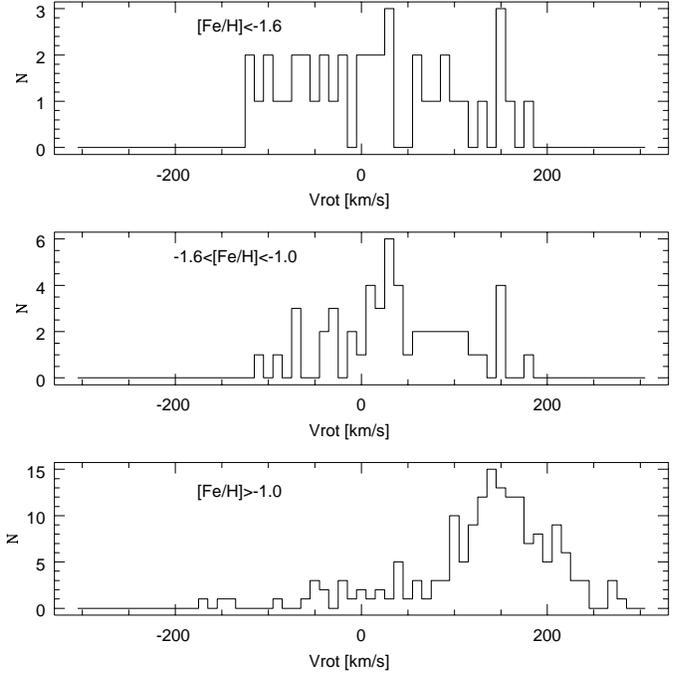}}
\caption{The same as Fig. {\ref{histo_ari}}, but for sample
CLLA-Tycho2.} \label{tychohist}
\end{figure}

The corresponding V$_\mathrm{rot}$-velocity distributions for the samples
CLLA-TYC2+HIP and CLLA-Tycho2 are shown in Figs.\ref{histo_ari} and
\ref{tychohist}, respectively. The first group, [Fe/H] $>-1$ dex,
represents what are obviously the thick disk stars. The third
group, [Fe/H] $<-1.6$ dex, consists of extreme metal-poor stars, dominated
by members of the halo population. The histograms of the samples
CLLA-TYC2+HIP and CLLA-Tycho2 can be fitted by Gaussian distributions.

The second group, $-1.6\le $ [Fe/H] $\le -1$ dex, shows a peculiar
kinematics. A Kolmogorov-Smirnov test, which avoids binning of the data, shows that the velocity distribution of the very metal-poor stars, [Fe/H] $\le -1.6$, in the combined sample is statistically different from the velocity distribution of the intermediate population, $-1.6 \le$ [Fe/H] $\le -1.0$. The maximum deviation of the normalized cumulative distribution between the two groups is $D=0.241$ and thus significantly larger than the critical value $D_{0.05}=0.196$ (Sachs, \cite{sachs}), which leads to a rejection of the hypothesis of the statistical similarity of the velocity distributions of the two groups. Similarly we have shown that the velocity distribution of the metal-poor stars is symmetric with respect to $V_{\mathrm{rot}}=0$ km s$^{-1}$ ($D=0.149$, $D_{0.05}=0.282$), whereas the velocity distribution of the intermediate group is asymmetric ($D=0.375$, $D_{0.05}=0.300$). Thus the intermediate group seems to represent a different population of halo stars.

On the other hand, the asymmetric drift ratio $\langle V \rangle / \sigma_{\mathrm{U}}^{2} \:=\:-0.007$ (CLLA-TYC2+HIP) is very similar to that of the thick disk stars, $\langle V \rangle / \sigma_{\mathrm{U}}^{2} \:=\:-0.008$. We conclude tentatively that the stars in the $-1.6 \le$ [Fe/H] $\le -1.0$ metallicity range represent a population of the dynamically hot metal-weak thick disk (MWTD).

\section{Summary and Discussion}

We have analyzed the kinematics of $742$ nearby metal-poor
subdwarf stars from the CLLA catalogue. The subdwarfs were cross-identified
with the TYC2+HIP and Tycho 2 Catalogues to find accurate
trigonometric parallaxes and proper motions. The accurate
Hipparcos parallaxes lead to an upward correction factor of 11\%
of the photometric distance scale of CLLA, and it was used to
correct the photometric distances of CLLA-Tycho2 stars.

The present analysis indicates that the solar neighborhood
subdwarf stars with [Fe/H] $<$-1.6 show halo kinematics
characterized by a radially elongated velocity ellipsoid and no
significant rotation. At a metallicity range of [Fe/H] $>-1$, our
samples show {\it{disklike}} kinematics. In the metallicity range $-1.6 \le$ [Fe/H] $\le -1.0$
we found a significant number of stars with kinematics not of halo stars but that of a dynamically not-metal-weak tail of the thick disk.

Chiba \& Beers (\cite{chiba00}) obtained a fraction of 30 \% of low-metallicity
stars in their nonkinematically selected solar neighborhood sample
with $-1.7 <$ [Fe/H] $\le$ -1.0, which is consistent with our result of $18 \%$.
Chiba \& Yoshii (\cite{chiba98}) analyzed the kinematics of red giants
and RR Lyrae stars in the solar neighborhood based on Hipparcos data.
They found in both red giant and RR Lyrae samples in the range $-1.6 <$ [Fe/H] $\le$ -1.0,
a fraction of $\sim$ 10 \% of stars in a population with a mean
velocity $\langle V_\mathrm{\phi} \rangle_\mathrm{disk} = 195$ km s$^{-1}$.

We must try to understand the implications of a significant
population of MWTD stars for theories of the formation and
evolution of the Galaxy. It should be kept in mind that, although
the MWTD population may contribute a large fraction of the
\emph{local} metal-poor stars, the (inner) halo population is
probably still dominated by the stars with
[Fe/H] $\leq-1.6$ within a few kiloparsecs of the Sun. Furthermore,
although we have emphasized the possible importance of the MWTD
population, it certainly still appears to be a minor constituent
of the entire thick disk population (Beers et al. {\cite{beers02}}).

If there is indeed a significant fraction of thick disk stars with
metal abundance $-1.6 \le$ [Fe/H] $\le -1$, as we have argued, this
finding may have significance for formation scenarios of the
Galaxy. An interesting scenario  for the origin of an MWTD
component may be the merging of satellite galaxies (Searle \&
Zinn, {\cite{searle}}), which are then accreted by a thin, fast
rotating, possibly metal-poor, Galactic disk (Quinn et al.
{\cite{quinn}}; Wyse {\cite{wyse}}). The dynamical heating of the
stellar component of this disk in connection with the accretion
process produces the thick disk. The kinematics of the halo
depends on the dynamics of the merging satellites, whereas the
kinematics of the thick disk are determined by the heating of the
rotating thin disk. Based on this merging picture of galaxy
formation, one might argue that the ``shredded satellite'' stars
retain a kinematic signature distinct from the thick disk part
that results from the heated thin disk. The kinematic trace of the
destroyed satellite, which is probably the origin of the MWTD
stars, would be visible in the mean orbital rotational velocity of
stars. Based on a spectroscopic survey of $\sim 2000$ F/G stars
$0.5-5$ kpc above the Galactic plane Gilmore et al.
(\cite{gilmore02}) determined a mean rotational velocity lag of
the shredded galaxies of $\sim 100$ km s$^{-1}$. The actual lag expected
from the shredded satellite depends predominantly on the initial
orbit and the amount of angular momentum transport in the merger
process and is not initially predictable in a specific case
(Gilmore et al. {\cite{gilmore02}}).

\begin{acknowledgements}
M.I.A acknowledges support for this work as part of a Ph.D thesis from
a DAAD scholarship.
\end{acknowledgements}

\end{document}